\documentclass[12pt,worldsci]{article}
\usepackage{graphicx}
\begin{document}
\title{GLUON PROPAGATOR IN THE\\
 LANDAU GAUGE FIXED\\ LATTICE QCD SIMULATION}
\author{Hideo Nakajima\\
{\em Dept. of Infor. Sci., Utsunomiya Univ., Utsunomiya 321-8585 Japan}\\
\vspace{0.3cm}
and \\
\vspace{0.3cm}
Sadataka Furui\\
{\em School of Sci. and Engr., Teikyo Univ., Utsunomiya 320-8551, Japan}}
\maketitle
\setlength{\baselineskip}{2.6ex}

\vspace{0.7cm}
\begin{abstract}
We measured the gluon propagator in the Landau gauge fixed QCD Langevin
simulation{\cite{lat97}} and studied the infra-red behaviour of the gluon 
propagator. 
The $(4^3 \times 8)$ lattice siulation was done for quenched
$\beta=3, 4$ and $5$ and unquenched $\beta=4, \kappa=0.1, 0.15$ and $0.2$,
using each 100 independent samples. The Landau 
gauge fixing was done by an extension of the Fourier acceleration method with
the condition $|{\rm div} A|<10^{-4}$, and the field A is related to the link variable
by $U={\rm exp} A$ instead of the usual U-linear definition. We confirmed gauge fixing
with smearing preconditioning{\cite{HdF}} works perfectly for the purpose of 
finding the global minimum of the squared norm of the gauge field when $\beta$ 
is large (e.g. $\beta=5$).
 Our simulation results suggests the  possibility
of a realization of the infra-red behaviour of the Gribov-Zwanziger theory.

\end{abstract}
%\vspace{0.7cm}

\section{Introduction}

In 1978 Gribov showed that the fixing of the divergence of the
gauge field in non-Abelian gauge theory does not uniquely fix its gauge.
The restriction of the gauge field such that its Faddeev-Popov
determinant is positive (Gribov region) is necessary yet insufficient
for excluding the ambiguity.

 The restriction to the fundamantal modular region, i.e. minimizing the norm of
the gauge field, cancels the infrared singularity of the perturbation theory 
and makes the gluon to have the complex mass.

\begin{eqnarray}
D_{\mu\nu}(k)&=&\displaystyle{{1\over n}}
\sum_{x={\bf x},t} e^{-ikx}Tr\langle A_\mu(x) A_\nu(0) \rangle\nonumber\\
&=&G(k^2)(\delta_{\mu\nu}-{k_\mu k_\nu\over k^2})\ \ .
\end{eqnarray}
\begin{equation}
 G(k^2)={k^2\over k^4+\kappa^4}={1\over 2}({1\over k^2+i\kappa^2}+
{1\over k^2-i\kappa^2})
\end{equation}
where $n=N^2-1$ is the dimension of the adjoint representation of the SU(N) group.

In order to measure the propagator on a lattice, it is necessary to avoid 
the Gribov copy.  We confirmed that it is possible by the smeared gauge 
fixing{\cite{HdF}}, when $\beta=5$ in the case of $4^3\times 8$.  
%\section{The Langevin simulation of lattice QCD and the Landau gauge fixing}

\section{The gluon propagator}

On the lattice we define an $SU(N)$ lattice action for $S=S_{gauge}+S_{fermi}$ in the
pseudo-fermion scheme. We measure
\begin{eqnarray}
D_T({\bf k},t)&=&{1\over 2 n}\sum_{j=1}^2 \sum_T Tr \langle \tilde
A_j({\bf k},T+t)\tilde A_j(-{\bf k},T)\rangle
\nonumber\\
&=&{1\over 2 n}\sum_{j=1}^2\sum_T \sum_{{\bf x}} e^{-i{\bf kx}}
Tr\langle A_j({\bf x},T+t) A_j({\bf 0},T)\rangle
\end{eqnarray}
where $\tilde A_j$ is the Fourier component of $A_j$, and
$j$ is summed over transverse polarization with respect to {\bf k}
.

 The gauge field on links is defined as
$e^{A_{x,\mu}}=U_{x,\mu}\ ,\  {\rm where}\ \ A_{x,\mu}^{\dag}=-A_{x,\mu}$
 The gauge transformation is
$e^{A^g_{x,\mu}} = g_{x}^{\dag}e^{A_{x,\mu}} g_{x+\mu}
\  ,\ \  \phi^g_x = g^{\dag}\phi_x\ \ $
where $g_x=e^{\epsilon_x}$ and $\epsilon$ is a traceless antihermitian 

The connected part of the propagator is defined by the normalized
Fourier transform
\begin{equation}
 \tilde A^a_\mu({\bf k},T)={1\over \sqrt{N_x^3}}\sum_x A^a_\mu({\bf x},T)
     e^{-i {\bf k}\cdot {\bf x}}
\end{equation}
as
\begin{eqnarray}
 G_{\mu\nu}({\bf k},t)_c&=& \sum_{T,a} \langle
    \tilde A^a_\mu({\bf k},T+t)\tilde A^a_\nu({\bf k},T)\rangle^*\nonumber\\
    &-&\sum_{T,a} \langle \tilde A^a_\mu({\bf k},T+t)\rangle 
     \langle\tilde A^a_\nu({\bf k},T)\rangle^*
\end{eqnarray}
where $\langle\tilde A^a_\mu({\bf k},T)\rangle$ specifies the expectation value of the gauge field 
in the Landau gauge. For this evaluation
it is necessary to fix the global gauge transformation.

We fix the global gauge such that the average over T of the zero-mode
of a sample: $\displaystyle\frac{1}{3 N_T}\sum_{T,i} \tilde A^m_i({\bf 0},T)\lambda_m, $ 
where $\lambda$ is the SU(3) Gell-Mann matrix, 
is diagonalized and the smallest eigenvalue appears at the top and the 
largest eigenvalue at the bottom. 
We observed that the 2nd term of 
$G_{ii}({\bf 0},t)_c$ is almost independent of t and that the value is several {\%} of
the 1st term of $G_{ii}({\bf 0},0)_c$. The value is larger for the unquenched case
as compared to the quenched case. The Fourier transform of $D_T({\bf 0},t)$
in the unquenched $4^3\times 8$ and in the quenched $8^3\times 16$ lattice simulation 
\cite{lat98} suggest the suppression of the infrared part
of the gluon propagator.
\begin{figure}[b]
\begin{centering}
\includegraphics[scale=.60]{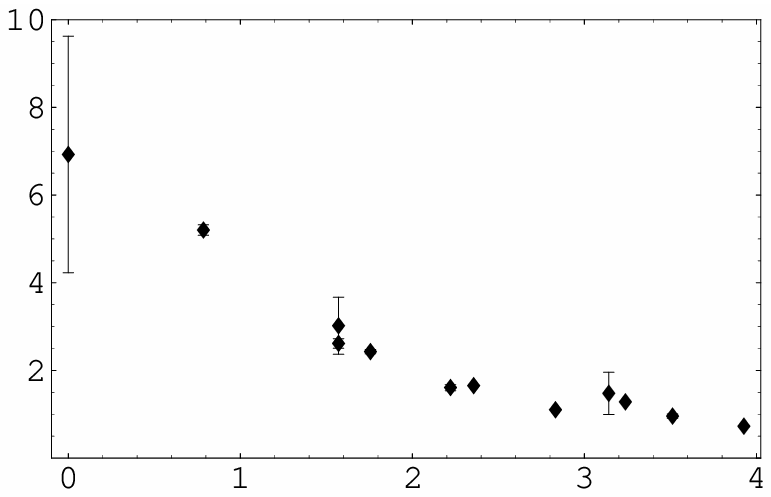}
\includegraphics[scale=.60]{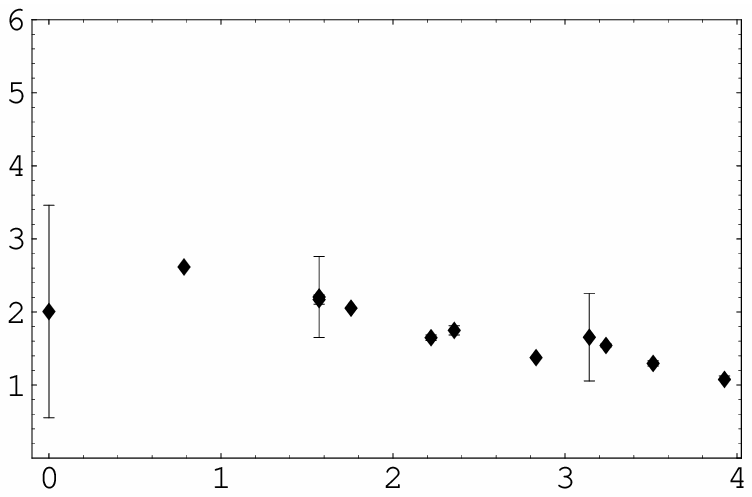}
  \caption[x]{The gluon propagator $D_T(p)$. Ordinates are the euclidean
4-momentum $p$ in the unit of $a^{-1}$. $\beta=4,\kappa=0$(left) and $\kappa=0.2$(right). $4^3\times 8$ lattice.}
  \label{fig:TheDt}
\end{centering}
\end{figure}
\vspace{0.3cm}
%
%\subsection{Parametrization of the gluon propagator}

In the Stingl's factorizing-denominator rational approximants
(FDRA) method{\cite{FS}}, the transverse gluon propagator is expressed in the 
2nd order as
\begin{equation}
D_T(p^2)^{[r=3]}={\rho\over p^2+u_+\Lambda^2}+{\tau\over p^2+v_+\Lambda^2}+c.c.
\end{equation}
We compared the lattice fourier transform of the above expression with 
$D_T({\bf k},t)$ and observed that in the confinement region there appears a
pair of complex zero near ${p^2=0}$ as is suggested by the eq (2).

\section{Discussion and outlook}

In the Landau gauge, the transverse gluon propagator in the infrared region is finite at $p^2=0$
\cite{Cu,LSWP}and the position of the zero of $D_T(p^2)$ approaches $p^2=0$ as the lattice size increases,
while the propagator of the Faddeev-Popov ghost which is measured in the quenched $8^3\times 16$
lattice shows the infrared singularity. These results support the possibility of a 
realization of the confinement mechanism of the Gribov-Zwanziger theory.

This work was supported by High Enery Accelerator Research Organization, as KEK Supercomputer
Project (Project No.98-34).

\end{document}